\documentclass{aa}
\usepackage{graphicx}
\def\ltsima{$\; \buildrel < \over \sim \;$}
\def\simlt{\lower.5ex\hbox{\ltsima}}
\def\gtsima{$\; \buildrel > \over \sim \;$}
\def\simgt{\lower.5ex\hbox{\gtsima}}
\def\cgs{{erg cm$^{-2}$ s$^{-1}$}}
\def\ergs{{erg s$^{-1}$}}
\def\cm2{{cm$^{-2}$}}

\def\xndof{{$\chi^{2}_{\rm \nu}$(dof)}}

\def\fhx{{$F_{2-10}$}}

\def\lhx{{$L_{2-10 keV}$}}

\def\p1{{Paper I}}

\def\fux{{$F_{20-30}$}}
\def\xmm{{\em XMM--Newton}}
\def\chandra{{\em Chandra}}
\def\beppo{{\em BeppoSAX}}

\def\nh{{N$_{\rm H}$}}

\def\nh{{N$_{\rm H}$}}

\def\epic{{\em EPIC}}
\def\mosuno{{\em MOS1}}
\def\mosdue{{\em MOS2}}
\def\pn{{\em PN}}
\def\mos{{\em MOS}}

\def\tre{{$\times$ 10$^{23}$ cm$^{-2}$}}
\def\f14{{10$^{-14}$}}
\def\f13{{10$^{-13}$}}
\def\ph{{photons cm$^{-2}$ s$^{-1}$}}

\def\feka{{Fe K$\alpha$}}
\def\fekb{{Fe K$\beta$}}
\def\iras{{IRAS 09104$+$4109}}
\begin{document}

\title{The XMM-Newton view of IRAS 09104$+$4109:\\
evidence for a changing-look Type 2 quasar?}
\author{E.~Piconcelli\inst{1}, F.~Fiore\inst{1}, F.~Nicastro\inst{1,2},
  S.~Mathur\inst{3}, M.~Brusa\inst{4}, A.~Comastri\inst{5},
  S.~Puccetti\inst{1,6}}

\titlerunning{XMM-Newton observation of IRAS 09104$+$4109}\authorrunning{E.~Piconcelli et al.}

\offprints{Enrico Piconcelli, \email{piconcelli@oa-roma.astro.it}}
\institute{Osservatorio Astronomico di Roma (INAF), Via Frascati 33, I--00040
  Monteporzio Catone (Roma), Italy\and Harvard-Smithsonian Center for
  Astrophysics, 60 Garden Street, Cambridge, MA 02138, USA\and Ohio State 
University, 140 West 18th Avenue, Columbus, OH 43210, USA\and Max Planck
Institut f\"ur Extraterrestrische Physik, Postfach 1312, D-85741 Garching bei
M\"unchen, Germany\and Osservatorio Astronomico di Bologna (INAF), via Ranzani
1, I--40127  Bologna, Italy\and
ASI Science Data Center, ESRIN, via G. Galilei, I--00044 Frascati, Italy}

\date{}

\abstract{} {We  report on a 14 ks \xmm~observation of the hyperluminous
infrared galaxy \iras, which harbors a Type 2 quasar in its nucleus. Our
analysis was aimed at studying the properties of the absorbing matter and the
Fe K complex at 6-7 keV in this source.}  {We analyzed the spectroscopic data
from the \pn~and the \mos~cameras in the 0.4--10 keV band.  We also used
an archival \beppo~1--50 keV observation of \iras~to investigate
possible variations of the quasar emission.}  
{The X-ray emission in the
\epic~band is dominated by the intra-cluster medium thermal emission. We
found that the quasar contributes $\sim$35\% of the total flux in the
2-10 keV band. Both a transmission- (through a Compton-thin absorber
with a Compton optical depth of $\tau_C$ $\sim$ 0.3, i.e. \nh~$\sim$5 $\times$ 10$^{23}$ \cm2) and 
a reflection-dominated ($\tau_C$ $>$ 1)
model provide an excellent fit to the quasar continuum emission.  However, the
value measured for the EW of \feka~emission line is only marginally
consistent with the  presence of a Compton-thick absorber in a
reflection-dominated scenario,
 which had been suggested by a previous, marginal (i.e. 2.5$\sigma$)
detection with the  hard X-ray (15--50 keV), non-imaging 
\beppo/PDS instrument. Moreover,  the value of luminosity in the 2--10 keV
  band  measured by the transmission-dominated model is fully consistent
with that expected on the basis of the bolometric
luminosity of \iras.
From the
analysis of the \xmm~data we  therefore suggest  the possibility that the absorber
along the line of sight to the nucleus of \iras~is
Compton-thin. 
Alternatively, the absorber column density  could have changed
from Compton-thick to -thin in the five years elapsed between the
observations. If this is the case, then \iras~is the first
``changing-look'' quasar ever detected.}  {}
% 5 {} token are mandatory

   \maketitle 
%
%________________________________________________________________

\section{Introduction}
\iras~is one the most powerful objects in the $z$\simlt0.5 Universe.  It is a
hyperluminous infrared cD galaxy of IR luminosity $>10^{13}L\odot$ at $z$ =
0.442, residing in the core of a rich cluster of galaxies (Kleinmann et
al. 1988).  Its optical spectrum shows only narrow emission lines, but broad
Balmer and Mg II emission lines were observed in the polarized light (Hines \&
Wills 1993; Tran et
al. 2000).  These pieces of evidence have led to the conclusion that a
dust-enshrouded Type 2 quasar lies in the nucleus of \iras.

The X-ray observations of \iras~collected so far  lend support to  this
suggestion.  Franceschini et al. (2000; F00) reported on the analysis of a
\beppo~observation of this source. They found that the X-ray spectrum below 10
keV is dominated by the intra-cluster medium (ICM) thermal ($k$T $\sim$ 5.5
keV) emission.  The detection of a weak signal in the 15--60 keV band with the
PDS instrument was interpreted by F00 as the  primary emission of the buried
quasar, which emerges from an absorbing  screen of \nh~\simgt~5 $\times$
10$^{24}$ \cm2.  They also reported on the marginal detection of a neutral
\feka~emission line with an equivalent width (EW) of $\sim$ 1--2 keV,
consistent with a reflection-dominated scenario for the quasar emission in the
2-10 keV band.  This evidence makes this source the best example of a
Compton-thick Type 2  quasar found to date.  Exploiting the high spatial
resolution of \chandra, Iwasawa et al. 2001 (I01 hereafter) were able to
analyze the spectral data of the central source  embedded in the extended ICM
emission.  Although limited by the low statistics ($\sim$ 200 counts in the
range 0.6-7 keV),  these data allowed to confirm the presence of a heavily
obscured quasar and, by the comparison with the \beppo~PDS flux, I01 concluded
that the \chandra~spectrum is also reflection-dominated.

\section{XMM--Newton observation and data reduction}

\iras~was observed with \xmm~(Jansen et al. 2001 and references therein) on
April 27, 2003 for $\sim$14 ks. The \epic~\pn~and \mos~observations were
carried out in the full frame mode using the Medium filter.  \xmm~data were
processed with SAS v6.5.  We used the  EPCHAIN and  EMCHAIN tasks for
processing the raw \pn~and \mos~data files, respectively, to generate
the relative linearized event files.  X--ray events corresponding to patterns
0--12(0--4)  for the \mos(\pn)~cameras were selected.  Hot and bad pixels were
removed.  The event lists were furthermore filtered to ignore periods of high
background flaring  according to the method presented in Piconcelli et
al. (2004) based on the cumulative distribution function of background
lightcurve count-rates.   Final net exposures of 10.8, 13.4, and 13.4 ks were
obtained  for \pn, \mosuno~and \mosdue, respectively.  The source photons were
extracted for the \pn(\mos) camera from a circular region with a radius of
37(40) arcsec, while the background counts were estimated from a larger
(i.e. $\sim$ 75 arcsec radius) source-free region on the same
chip. Appropriate response and ancillary files for all the \epic~cameras were
created using respectively RMFGEN and ARFGEN tasks in the SAS. Combined {\em
MOS1}$+${\em MOS2}  spectrum and response matrix were created.
%%%%%%%%%%%%%%%%%%
%%%%%%%%%%%%%%%%%%%%%%%%%%%%%%%%%%%%%%%% flux 0.5 7 2.0334E-12
\begin{table*}
\caption{Best-fit spectral parameters of the \epic~spectrum. 
See Sect.~\ref{s:spex} for details. Column: (1) spectral model; (2) 
temperature of the ICM ``cool core''  component (keV); (3) 
ICM metallicity; (4)  2--10 keV flux of the quasar component (10$^{-13}$
\cgs); (5)  2--10 keV luminosity of the quasar component (10$^{44}$ \ergs);
(6)  column density of the absorber (10$^{23}$ \cm2); (7)  energy of the
\feka~line (keV); (8)  intensity of the \feka~line (10$^{-6}$
ph/cm$^{2}$/s); (9)  EW of the \feka~line (eV); (10)  reduced $\chi^2$ and number
of degrees of freedom.}
\label{tab:fit}
\begin{center}
\begin{tabular}{ccccccccccc}
\hline \multicolumn{1}{c} {Model$^a$}& \multicolumn{1}{c} {$k$T}&
\multicolumn{1}{c} {$Z/Z_\odot$}& \multicolumn{1}{c} {$F_{2-10}$}&
\multicolumn{1}{c} {$L_{2-10}$}& \multicolumn{1}{c} {N$_{\rm H}$}&
\multicolumn{1}{c} {E$_{K\alpha}$}& \multicolumn{1}{c} {I$_{K\alpha}$}&
\multicolumn{1}{c} {EW$_{K\alpha}$}&
\multicolumn{1}{c} {\xndof}\\ 
(1)&(2)&(3)&(4)&(5)&(6)&(7)&(8)&(9)&(10)\\
\hline\\ (A)&3.9$^{+0.8}_{-0.9}$&0.47$^{+0.10}_{-0.10}$&4.68&7.95
&4.8$^{+2.6}_{-1.5}$&6.38$^{+0.06}_{-0.06}$&8.7$^{+6.2}_{-4.8}$&390$^{+380,
b}_{-212}$&0.94(219)\\
(B)&3.6$^{+0.8}_{-0.7}$&0.48$^{+0.11}_{-0.10}$&5.57&$>$2.05
&$-$&6.38$^{+0.05}_{-0.05}$&4.0$^{+2.1}_{-1.9}$&402$^{+212, c}_{-193}$&0.96(220)\\ \hline
\end{tabular}
\end{center}
$^a$ (A) {\it transmission} model; (B) {\it reflection} model;  $^b$
absorption-corrected line against absorption-corrected continuum; $^c$ with respect to
the pure reflection component.
\label{tab:fit}
\end{table*}
%%%%%%%%%%%%%%%%%%%%%%%%%%%%%%%%%%%%%%%%%%%%% $^{+ }_{- }$
%%%%%%%%%%%%%%%%%%%%%%%%% $^{+ }_{- }$

Light-curves in the 0.2--2 keV and 2--15 keV band were  extracted.  Both
light-curves are consistent with a constant flux level over the whole
\xmm~exposure.
\section{Spectral modelling}
\label{s:spex}

Both  \pn~and \mos~spectra were grouped to have a minimum of 20 counts per bin
to allow the use of  $\chi^{2}$ minimization technique and fitted simultaneously.
Given the current calibration uncertainties,  we discarded events below 0.4
keV and above 10 keV.  All fits were performed using the XSPEC package (v11.3)
and included the Galactic column density value of \nh=1.81 $\times$ 10$^{20}$
\cm2 (Murphy et al. 1996).  Best-fit parameter values are given in the source
frame, unless otherwise specified.  The quoted errors
on the model parameters correspond to a 90\% confidence level for one
interesting parameter ($\Delta\chi^2$ = 2.71; Avni 1976). A cosmology with ($\Omega_{\rm
M}$,$\Omega_{\rm \Lambda}$) = (0.3,0.7) and a $H_{\rm 0}$ = 70 km s$^{-1}$
Mpc$^{-1}$ is assumed throughout.

%%%%%%%%%%%%%%%%%%%
%Spectral analysis%
%%%%%%%%%%%%%%%%%%%

The high-resolution (\simlt~1 arcsec) \chandra~observation of \iras~presented by I01 has
definitively established that the emission from the ICM dominates 
the X-ray spectrum below 10 keV. Furthermore \iras~is a massive
  cooling-flow cluster (Fabian \& Crawford 1995) 
showing  a large radial temperature gradient, from $k$T $\approx$ 3.3 keV in the
bright, cool core up to $k$T $\approx$ 7.8 keV at a distance of 200 kpc,
i.e. 36 arcsec (I01).
For this reason, we
initially fitted the combined \pn$+$\mos~spectrum with a model
consisting of two thermal plasma emission components (i.e. {\tt MEKAL} in
{\tt XSPEC}) plus an absorbed power law. 
We fixed the temperature of one thermal component to 7.8 keV
in order not to underestimate the total ICM emission in the hard X-ray band.
%The temperature measured
%for the ICM by a fit with a single thermal-emission component
%would indeed be an average value, dominated by the temperature of the cool core
Since the photon index of the
power law was loosely constrained ($\Gamma$ $\approx$1.4$\pm$1) we fixed
it to 1.8, which is the average value typically observed for radio-quiet
quasars (e.g.  Piconcelli et al. 2005).

%======================================================
\begin{figure*}
\begin{center}
\includegraphics[width=5.cm,height=7.cm,angle=-90]{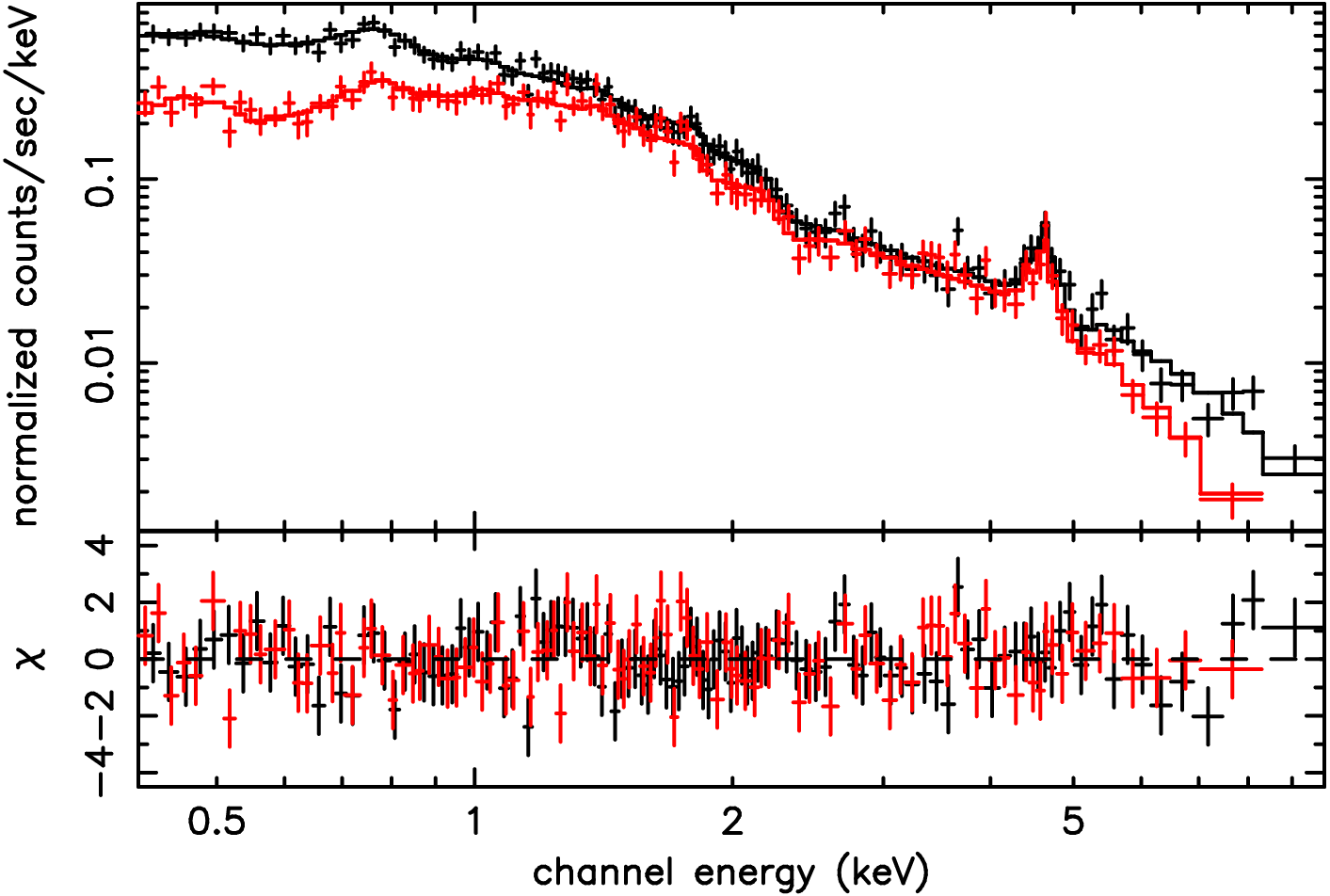}
\hspace{0.5cm}\includegraphics[width=5.cm,height=7.cm,angle=-90]{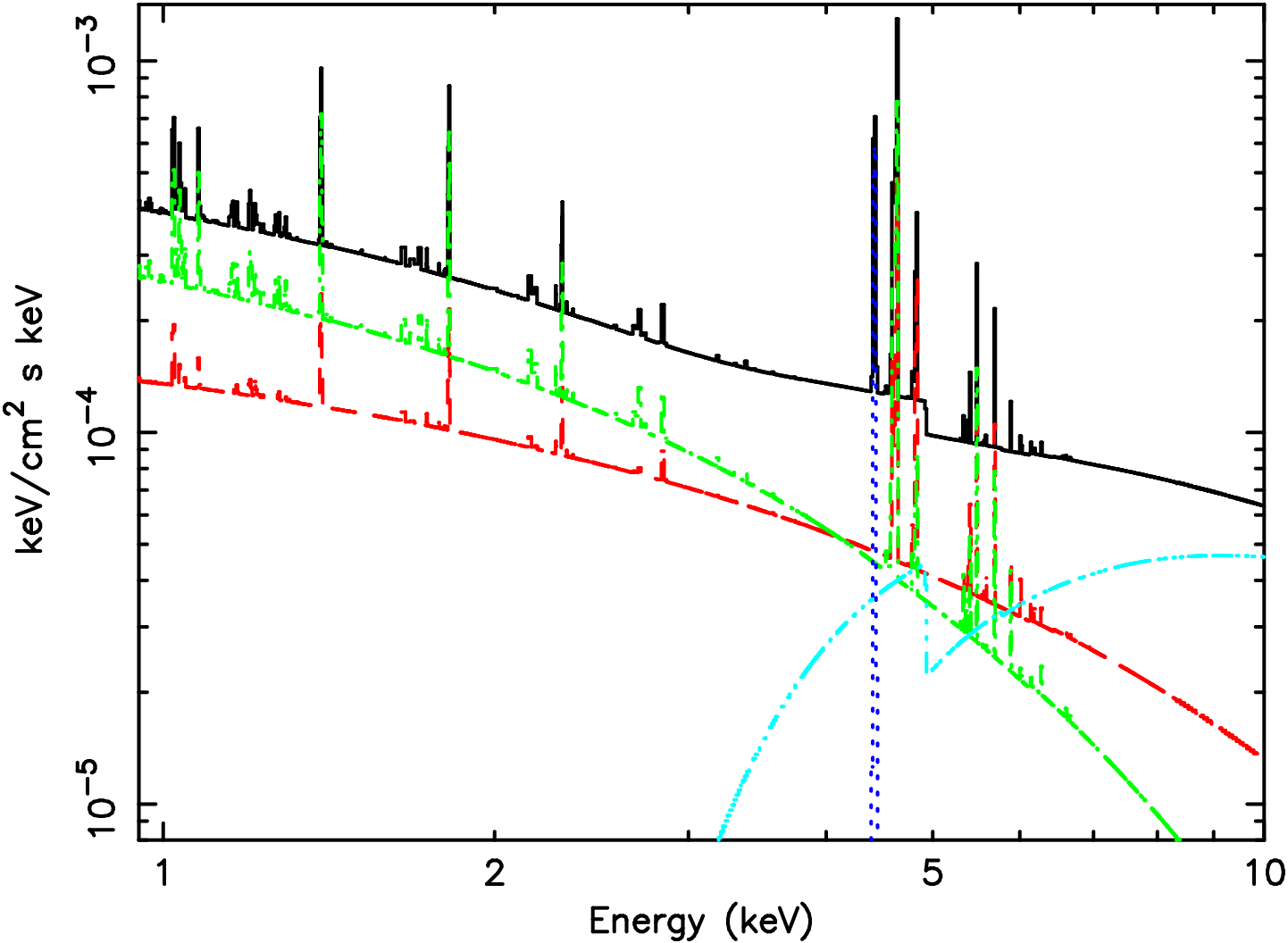}
\caption{(a) {\it Left:} \xmm~\pn~(top) and \mos~(bottom) spectra of
  \iras~when the {\it transmission} model is applied.  The lower panel shows
  the deviations of the observed data from the model in unit of standard
  deviations.   (b) {\it Right:} Best-fit model for the transmission scenario.
  The different spectral components are also plotted (e.g. Section~\ref{s:spex}
  for further details).}
\label{fig:trans}
\end{center}
\end{figure*}
%==
%======================================================
This model ({\it transmission} model hereafter) gave an excellent description
of the \epic~data with a \xndof~=0.94(219) (see Fig.~\ref{fig:trans}). The
best-fit values of the spectral parameters are listed in Table 1. The value
of $\sim$ 4.8 $\times$ 10$^{23}$ \cm2~found for the column density 
  translates in a Compton optical depth  $\tau_C$ $\sim$ 0.3
of the absorbing screen. The value of temperature ($k$T =
3.9$^{+0.8}_{-0.9}$ keV) in the cluster core and metallicity ($Z/Z_\odot$ =
0.47$^{+0.10}_{-0.10}$) of the ICM are consistent with I01.
%======================================================
\begin{figure*}
\begin{center}
\includegraphics[width=5cm,height=7.0cm,angle=-90]{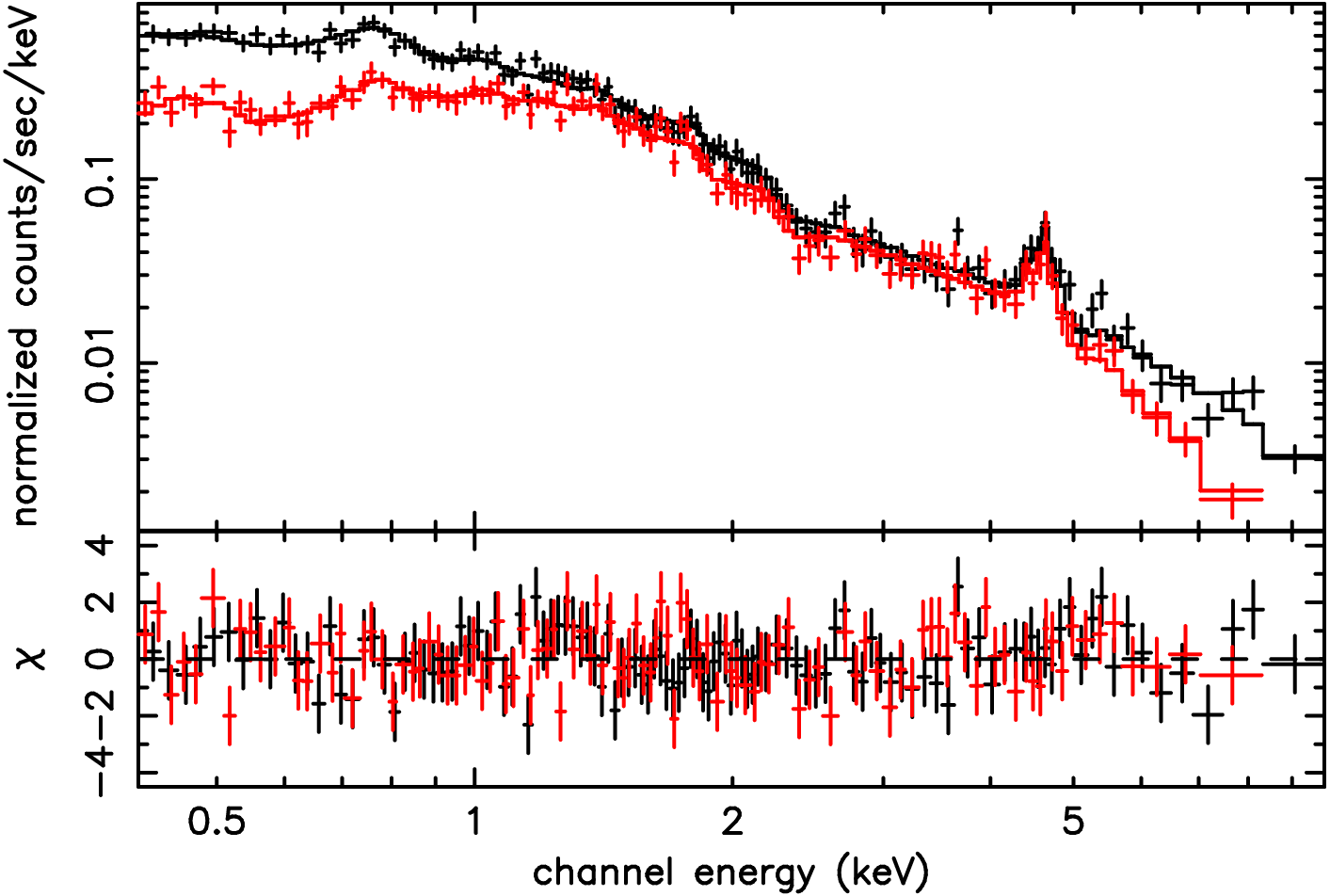}
\hspace{0.5cm}\includegraphics[width=5cm,height=7.0cm,angle=-90]{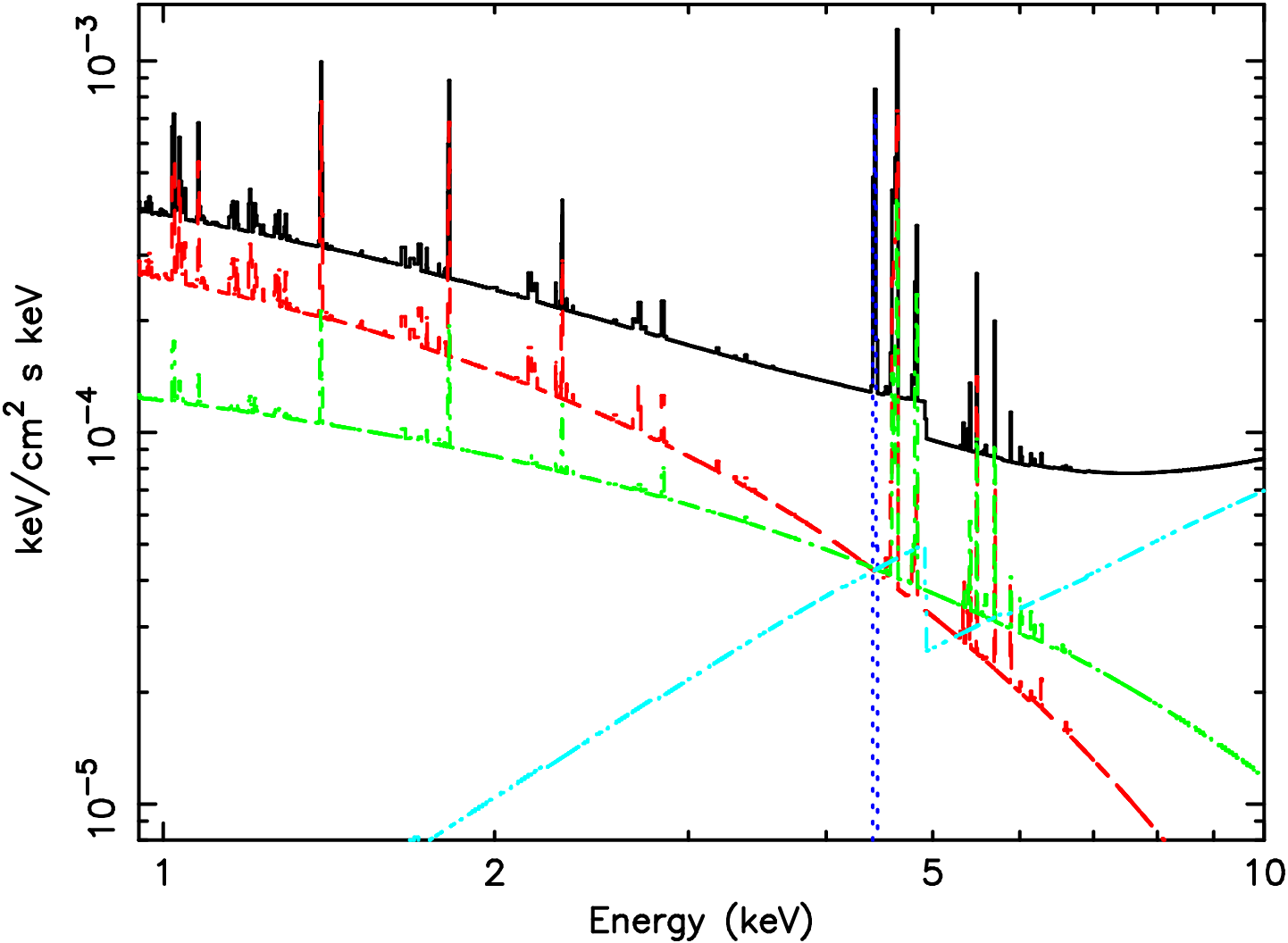}
\caption{ The same as Fig.~\ref{fig:trans} for the {\it reflection} model.}
\label{fig:refl}
\end{center}
\end{figure*}
%==
%======================================================

Works based on \beppo~and \chandra~observations favored an interpretation of
the spectrum of \iras~below 10 keV in terms of reflection-dominated emission.
We therefore replaced the absorbed power law in the {\it transmission} model
with a   Compton reflection component from neutral matter (i.e. {\tt PEXRAV}
model in {\tt XSPEC}).  For this spectral component, which is due to the
reprocessing of the emission from the obscured primary X-ray source, we
assumed a $\Gamma$ = 1.8 for the photon index of the incident power law, along
with an inclination angle of $i$ = 50 deg (Tran et al. 2000) and solar
metallicity for the reflector.  This model ({\it reflection} model hereafter)
yielded an equally good fit to the \xmm~spectrum with a final \xndof =
0.96(220) (see Table 1 and Fig.~\ref{fig:refl}).
%The measured flux of the quasar component at 2--10 keV was 
%\fhx~$\sim$ 6.4 $\times$ 10$^{-13}$ \cgs.
According to both spectral models, the contribution of the quasar component to
the total flux in the 2--10 keV band is $\approx$30--35\%.

As shown in Figs.~\ref{fig:trans}a and \ref{fig:refl}a, 
there is a prominent line-like emission feature at 4-5 keV (observer-frame)
which is broader than the instrumental resolution at this energy
 and most likely due to a blend of lines associated with Fe K emission. 
This complex is partly accounted for by the strong FeXXV K$\alpha$ emission line at 6.7
keV and the FeXXVI K$\alpha$ emission line at 6.97 keV (which should be likely blended with 
a weak \fekb~line
at 7.06 keV due to the reprocessing of the quasar continuum) from the
two-temperature ICM. However, significant (at $>$99\% confidence level)
positive residuals are still present. We modelled this excess with an
unresolved Gaussian emission line at  6.38$^{+0.06}_{-0.06}$ keV. This energy
is consistent 
with a range of ionization states from
FeI to FeXVI (Kallman et al. 2004), as typically observed in quasars
(Jimenez-Bailon et al. 2005).
In  the ``reflection-dominated'' scenario  we
measure an equivalent width of the \feka~line at $\sim$6.4 keV of
EW$_{K\alpha}$ = 402$^{+212}_{-193}$ eV (calculated
with respect to the Compton reflection component). 
For the {\it transmission} model (i.e. assuming that the line and the continuum are both
absorbed) we derive EW$_{K\alpha}$ = 390$^{+380}_{-212}$ eV (see Table 1).

%\begin{figure}
%\begin{center}
%\includegraphics[width=5.cm,height=7cm,angle=-90]{righette.ps} \hspace{0.5cm}
%\caption{Close-up of the \pn~unfolded spectrum in the
%  observer-frame \feka~ emission band fitted with  the {\it reflection}
%model. Three emission lines clearly emerge from the underlying continuum
%(e.g. Sect.~\ref{s:spex} for details).}
%\label{f:righe}
%\end{center}
%\end{figure}

\subsection{A comparison with broadband 1--50 keV \beppo~data}
\label{sec:sax}

\begin{figure*}
\begin{center}
\includegraphics[width=5.cm,height=7.cm,angle=-90]{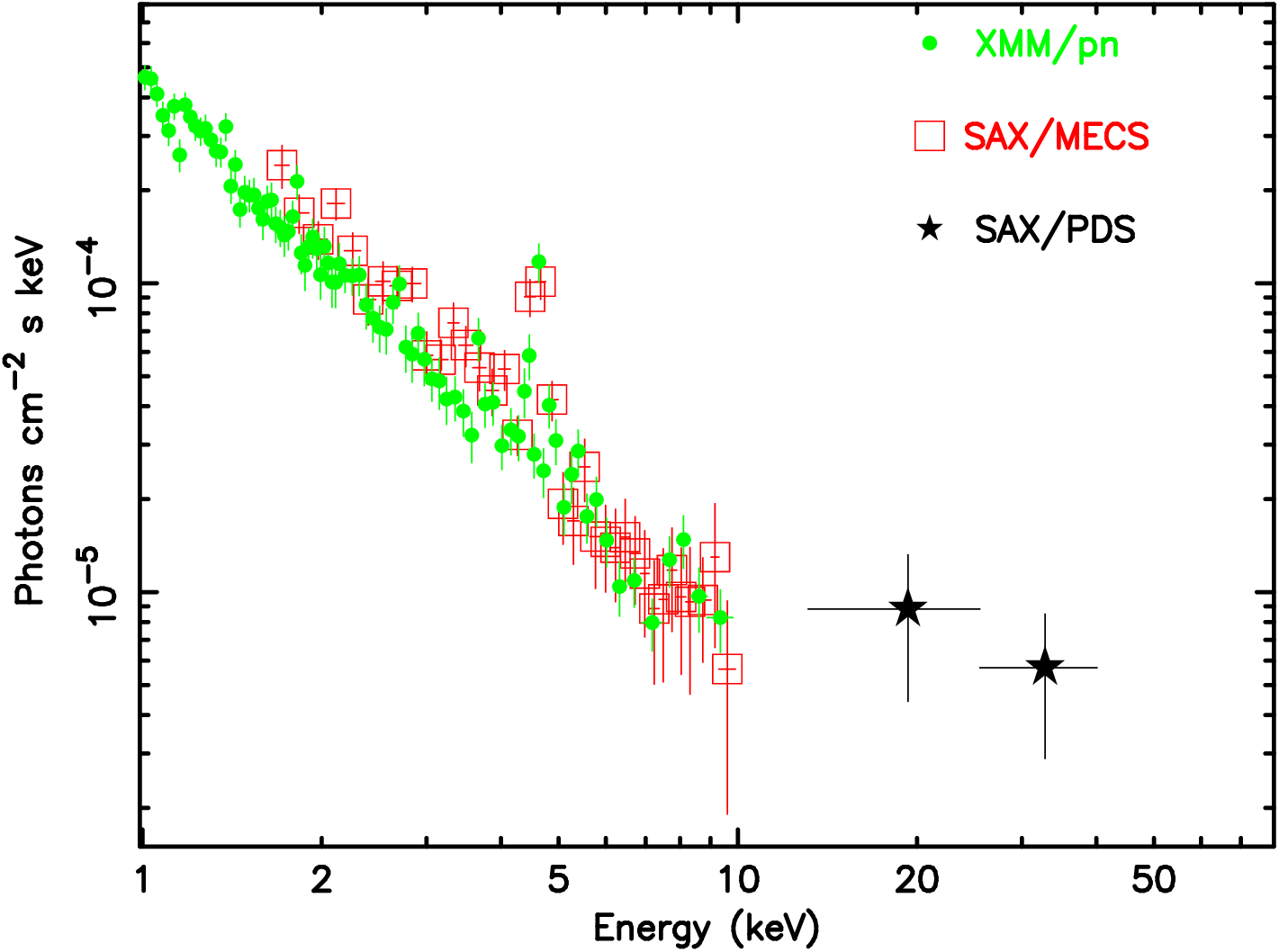}
\hspace{0.5cm} \includegraphics[width=5.cm,height=7.cm,angle=-90]{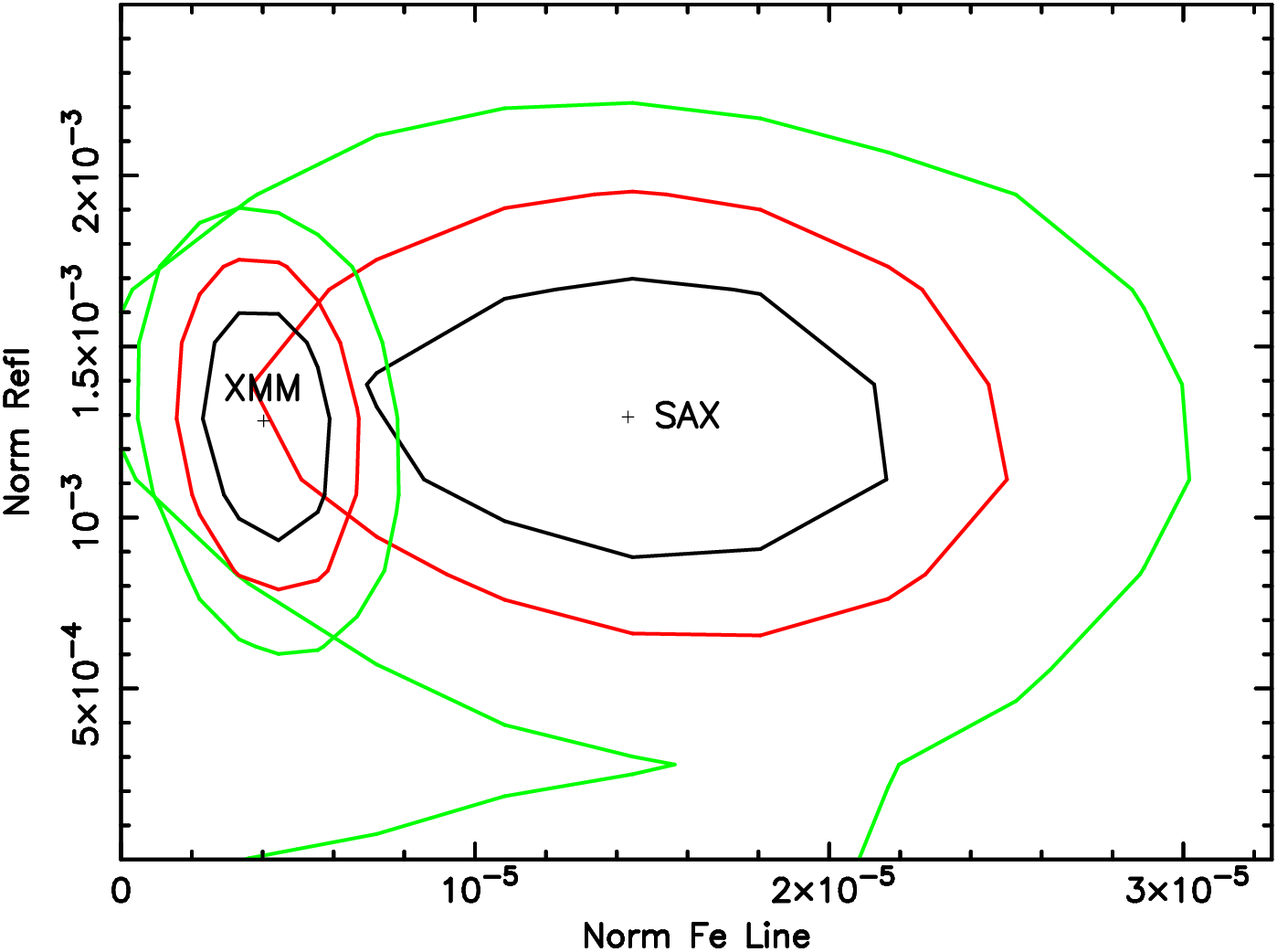}
%\hspace{0.5cm}
%\includegraphics[width=5.5cm,height=7.5cm,angle=-90]{modref_j.ps}
\caption{(a) {\it Left:} Unfolded spectra for the 2003 \xmm~and 1998
  \beppo~observations.
(b) {\it Right:}  Confidence contour plot showing the
intensity of the \feka~emission line (in units of ph/cm$^{2}$/s) against the
intensity (i.e. photon flux at 1 keV) of the reflection continuum (in units of
ph/keV/cm$^{2}$/s) obtained using the \xmm~and \beppo~(MECS$+$PDS) data.
 The contours are at 68\%, 90\% and 99\% confidence levels for two interesting parameters.}
\label{fig:uf}
\end{center}
\end{figure*}

To investigate on the possible  year-timescale variability of the overall continuum
spectral shape, we plot in Fig.~\ref{fig:uf}a  the unfolded 2003 \pn~and
1998 \beppo~MECS+PDS spectra.  These data have been unfolded through the instrument
response with respect to the best-fit model found by F00 (i.e. ICM thermal
emission component $+$ absorbed powerlaw $+$ reflection $+$ narrow Gaussian
line at 6.4 keV; model {\it sax98} hereafter).  As expected,
the 1--10 keV \xmm~and \beppo~spectra have similar shape being dominated by
the ICM emission.  The different normalizations of the two spectra can be
ascribed to the different source extraction regions. In particular, a fraction of ICM
emission from the outskirt region of the cluster spreads outside the
\pn/\mos~extraction radius.
  Interestingly,  the PDS data above 10 keV appear to lie slightly
above the extrapolation of the MECS/\pn~continuum level.
We calculated a 20--30 keV flux level of \fux~$\sim$ 6.1 $\times$ 10$^{-13}$
and $\sim$ 1.3$\times$ 10$^{-12}$ \cgs~for the  {\it transmission}\footnote{In
the case of this model, we also added to the model a Compton reflection
component with $R$ = 1 (where $R$ is the solid angle in units of 2$\pi$
subtended by the reflecting material) as commonly observed in the 10--50 keV
AGN spectra (e.g. Risaliti 2002; Reeves et al. 2006).}  and  {\it reflection}
model, respectively.  The  20--30 keV flux of \iras~measured by \beppo~is
\fux~= 2.55$^{+1.90}_{-1.56}$  $\times$ 10$^{-12}$ \cgs, whereby the extrapolated \xmm~flux in 
this band is fainter by a factor of $\sim$1.6--7.3, 
if the {\it transmission} model is assumed.
On the other hand, 
the 20--30 keV flux estimated by {\it reflection} model is  consistent with
the \beppo~value.

%======================================================
%==================================================

Fig.~\ref{fig:uf}b shows the iso-$\chi^2$ contour plot of the intensity of the
\feka~emission line at 6.4 keV (I$_{K\alpha}$) versus the intensity of the
reflection continuum for the 1998 \beppo~and 2003 \xmm~observations.  The
\xmm~values were derived by the  {\it  reflection} model, while for the
\beppo~data we employed the {\it sax98} model (temperature and abundance of
the ICM were permitted to vary within the 90\% confidence interval measured
for these parameters with \xmm, e.g. Table 1).
These measurements taken at different epochs are consistent within
3$\sigma$ errors, as expected if Compton hump and cold iron emission
arise from distant material. Furthermore, the \xmm~measurements are also
compatible with the values of I$_{K\alpha}$ = 7.2$^{+6.1}_{-3.3}$
$\times$ 10$^{-6}$ \ph~and EW$_{K\alpha}$ = 1.1$^{+0.9}_{-0.5}$ keV
reported by I01 on the basis of a 1999 \chandra~observation of
\iras. The mean values of EW$_{K\alpha}$ were significantly
larger, but the errors were also very large in the low S/N \beppo~and
\chandra~spectra.
\section{Discussion}
\label{s:discussion}

The \xmm~observation presented here has confirmed the presence of a heavily absorbed
quasar in the  nucleus of the hyperluminous infrared galaxy \iras.  The \epic~energy range is
dominated by the thermal  emission of the ICM and we
estimated that the quasar contributes  approximately $\sim$30--35\% of the total
2--10 keV flux.  We were able to accurately study the Fe K
complex at $\sim$6--7 keV in this source,  modelling it with three distinct narrow emission
lines from  FeI--XVI K$\alpha$, FeXXV K$\alpha$ and  FeXXVI
K$\alpha$ transition,  respectively. The highly ionized
Fe K lines originate in the ICM, while the line from cold iron is likely due
to reprocessing of the quasar continuum off the circumnuclear environment.  The
better quality of \epic~data has allowed to overcome the problems of limited
spectral resolution and statistics of \beppo~data (e.g. F00),  which prevented
a correct estimate of the contribution from the different ions to the Fe
emission complex.

The quasar emission can be equally well fitted by either a {\it
transmission} or a {\it reflection} model. The former implies a
Compton-thin absorber with \nh~$\approx$ 5 \tre, while the latter
suggests a scenario where the primary X-ray continuum is blocked by a
Compton-thick ($\tau_C$ \simgt~1, i.e. \nh\simgt$\sigma^{-1}_t$ $\approx$1.6 $\times$
10$^{24}$ \cm2) obscuring screen and the emission observed by \xmm~is
due to indirect radiation scattered into the line of sight (it is
generally assumed that the absorber and the reflector are the same
material, i.e. the {\it torus}, e.g. Matt et al. 1996; Molendi et al.
2003).  Using the present data these models are statistically
indistinguishable.  However, an EW$_{K\alpha}$ $\approx$ 400$\pm$200 eV
measured assuming the {\it reflection} model is significantly lower than
expected for a truly ``reflection-dominated'' spectrum with [Fe/H] = 0.  In fact, the
most prominent feature in the 2-10 keV spectrum of a heavily obscured AGN, as
in the case of \iras~(2 $\times$ 10$^{24}$ \simlt~\nh~\simlt~10$^{25}$ \cm2, e.g. I01), 
is a strong \feka~emission line with an EW$_{K\alpha}$ \simgt~1 keV
(Levenson et al. 2002; Guainazzi et al. 2005).  
This solid observational evidence matches well
with the EW value predicted by the theoretical calculations (e.g. Leahy \&
Creighton 1993; Ghisellini et
al. 1994; Matt et al. 1996).  
%Thus, the {\it reflection} model, though statistically
%acceptable, does not seem to be physically plausible in \iras~because of  
%the weakness of the \feka~emission line. 
 However, given the uncertainty in the reflection continuum flux 
due to the presence of the diffuse cluster emission, we also conservatively estimated the
EW$_{K\alpha}$ by using the lowest(highest) value in the 90\% confidence interval measured for the 
normalization of the reflection component(\feka~emission line) and vice versa.
We obtained a range of EW$_{K\alpha}$ values  spanning from $\sim$165 to 870
eV, which is 
marginally consistent with the EW$_{K\alpha}$ $\approx$ 1 keV expected for a
reflection-dominated scenario.
 As the continuum flux estimated for the obscured quasar with 
the {\it reflection} model (\fhx~= 4.68$^{+1.63}_{-1.08}$ $\times$ 10$^{-13}$ \cgs)
  is similar to the value measured with the spatially resolved \chandra~spectrum (\fhx~$\approx$ 3.9
  $\times$ 10$^{-13}$ \cgs, e.g. I01), the marginal
  discrepancy in the  EW$_{K\alpha}$ values between the two observations
   arises from the different best-fit line flux intensity.\\

On the
other hand, the EW$_{K\alpha}$ = 390$^{+380}_{-212}$ eV (note the large
error bars) found in the case of the {\it transmission} model is in agreement with the
expected EW value of a \feka~line transmitted through an absorbing screen with
\nh~$\sim$5 \tre~is ($\sim$200--400 eV; e.g. Awaki et al. 1991; Leahy \&
Creighton 1993; Ghisellini et al. 1994).  
A scenario with a Compton-thin absorber along
the line of sight to the nucleus of \iras~is therefore physically plausible. 
 Nonetheless, the flux difference by a factor
of $\sim$1.6--7.3 between  the 2003 \xmm~and 1998 \beppo~observation (a time interval
of 3.5 yr at the source frame after the
  correction for the time dilation due to the $z$ = 0.442) in the 20-30 keV
band is puzzling. 
In fact, given the observed power spectral densities of low 
black hole mass (i.e. $M_{BH}$ $\approx$ 10$^7$ M$_\odot$) Seyfert galaxies
(Markowitz et al. 2003) and scaling linearly with the black hole mass
(assuming $M_{BH}$ =
2.4 $\times$ 10$^9$ M$_\odot$ for the nuclear black hole
 in \iras\footnote {We estimated a mass of 2.4 $\times$ 10$^9$
M$_\odot$ for the black hole  in the nucleus of \iras~using the measurement of
the MgII line width
from Hines \& Wills (1993), i.e. FWHM = 10,000 km/s, and the formula in
Willott et al. (2003).}),
flux variations of a factor of \simgt~2 in 3.5 yr should be considered unlikely
(e.g. Fiore et al. 1998).
However, the major problem in the interpretation of this mismatch concerns the
accuracy of \beppo/PDS measurement.  According to our analysis, the
signal-to-noise ratio of these data is poor (only $\sim$2.5$\sigma$ between 15
and 50 keV).
This implies that any conclusion based only on the PDS data about the
Compton-thick or Compton-thin nature of  the absorber in the nucleus of
\iras~could be misleading.  The addition of a primary continuum power-law
component  modified by a Compton-thick absorber (i.e. similarly to the
best-fit model  proposed by F00 for the \beppo~broadband spectrum) might
easily account for the above flux discrepancy, but, as stressed above,   a
scenario with a Compton-thick absorber
for the \epic~spectrum is at odds  with the
inferred quite small value (i.e. $<$ 1 keV) of the \feka~EW.

Given the  large 1.3$^\circ$ FWHM PDS field of view (FOV),
PDS data  might be affected by  the contamination of very hard X-ray
sources possibly located outside the \xmm~and \chandra~15 arcmin radius FOVs.
 If a power-law source with $\Gamma$ = 1.8 is assumed, the 20--30 keV flux 
measured for
the PDS source (\fux~$\sim$2.55 $\times$ 10$^{-12}$ \cgs) is translated 
to \fhx $\sim$ 7.5 $\times$ 10$^{-12}$ \cgs.
According to the RXTE all-sky slew survey log $N$ -- log $S$ function in the
2--10 keV band (Revnivtsev et al. 2004), at this flux level 0.025 sources are 
expected in a 1.5 deg$^2$ area, which translates into a probability of 2.4\%.
Moreover, using the NED and SIMBAD catalogs we found a probable contaminating source
in the H$_{\rm 2}$O maser galaxy NGC 2782 ($z$ = 0.008), which 
is located at $\sim$50 arcmin away
from \iras. The \chandra~spectrum of this source (with a nuclear X-ray flux of a few
times 10$^{-13}$ \cgs) strongly suggests the
presence of a heavily (\nh~$>$ 10$^{24}$ \cm2) absorbed AGN  (Zhang et
al. 2006), whereby it is likely that NGC 2782 provides a sizable contribution
to the  20--30 keV flux measured with the PDS. Assuming a 
pure reflection model, we estimated a  \fux~$\sim$6 $\times$ 10$^{-13}$ \cgs,
which must be considered a lower limit of the  20--30 keV emission from 
NGC 2782, because of the likely presence of the nuclear continuum emerging
after transmission through the absorber.\\

Alternatively, \iras~could represent the first example of a
``changing-look'' quasar (Guainazzi et al. 2002; Matt et al. 2003 and reference therein) ever
detected given that the \xmm~data are better explained by a
transmission-dominated model with \nh~$\sim$5 \tre~while, if the PDS
emission is entirely due to \iras, the \beppo~data are consistent with a
reflection-dominated spectral state. In fact, similar spectral
transitions from a Compton-thick to a Compton-thin state (or vice-versa)
have been observed, but only in Seyfert-like AGNs so far. This scenario
implies a dramatic decrease (a factor of \simgt5--10) in the line-of-sight
absorbing column density during a timescale of 5 years and, in turn,
suggests the presence of a largely inhomogeneous obscuring circumnuclear
gas (Elvis et al. 2004; Elitzur \& Shlosman 2006).
 We estimated the line-of-sight crossing-time of an obscuring cloud in Keplerian motion
around the central black hole of 2.4 $\times$ 10$^9$ M$_\odot$ to explain
the possible transition from a reflection-dominated to a transmission-dominated spectrum.
We assumed a scenario similar to that described for NGC 1365 in Risaliti et al. (2007) (or NGC 3227,
e.g. Lamer et al. 2003) where they found that the Compton-thick obscuring material
responsible of the spectral transition is located in
the broad line region (BLR) and the size of the X-ray emitting region is \simlt~100 R$_G$.
We estimated a distance of the BLR $R_{BLR}$ = 0.14 pc using the empirical
relation $R_{BLR}$--$M_{BH}$ in Kaspi et al. (2000).
The redshift-corrected crossing-time (Guainazzi et al. 2002) of a
Keplerian cloud covering a region of size 100 R$_G$ around a black hole of  
2.4 $\times$ 10$^9$ M$_\odot$ is $\sim$1.7 yr.
Such a value is therefore consistent with the  3.5 yr (source
frame) elapsed between the \beppo~and \xmm~observation.

Finally, we calculated a ratio $r_{X,bol}$ $\equiv$ \lhx/L$_{\rm bol}$ = 0.016 using the
2--10 keV luminosity
of \lhx~= 7.95 $\times$ 10$^{44}$ \ergs~measured
for the transmission-scenario (see Table 1), and a bolometric luminosity of
L$_{\rm bol}$($\approx$ L$_{\rm IR}$) = 4.7 $\times$ 10$^{46}$ \ergs,
 which is largely dominated by the obscured quasar (e.g. Hines et
  al. 1999; Spoon et al. 2007).
We also derived the value of $r_{X,bol}$ expected for \iras~using 
the value of $\nu$$l_{\nu}$(2500\AA)/L$_{\rm bol}$  typical for quasars reported in 
Elvis et al. (1994), correcting the L$_{\rm bol}$ value by a
factor of 1/3 as suggested by Fabian \&
Iwasawa (1999) in order not to  count twice the UV emission, and
 the spectral index $\alpha_{OX}$ between 2500\AA~and 2 keV,
defined as  $\alpha_{OX}$ = $-$0.384[$l_{\nu}$(2 keV)/$l_{\nu}$(2500\AA)] (e.g.
Tananbaum et al. 1979).
In particular, we used the relation $\alpha_{OX}$ = 0.137{\it
  log}($l_{\nu}$(2500\AA)) - 2.638 reported by  Steffen
et al (2006).
We converted from the monochromatic value of $\nu$$l_{\nu}$(2 keV) to the
\lhx~value by multiplying by a factor of 1.61, applying a photon index of
$\Gamma$ = 2. 
We obtained that $r_{X,bol}$ $\equiv$ \lhx/L$_{\rm bol}$ = 0.043 $\times$ L$_{\rm bol,45}$$^{-0.357}$
(with L$_{\rm bol,45}$=L$_{\rm bol}$ /10$^{45}$ \ergs).
This implies that the expected value of $r_{X,bol}$ for \iras~is 0.011, which
is close to the value of $r_{X,bol}$ = 0.016 measured using the  \lhx~derived
for the transmission scenario.
Since the observed luminosity of \lhx~= 2.05 $\times$ 10$^{44}$ \ergs~in the  {\it reflection}
model should be just few percent of the intrinsic one (Comastri 2004; I01),
this result lends further support to the hypothesis of a
Compton-thin absorber along the line of sight to the nucleus of \iras.\\

\section{Summary}
\label{s:summary}

The  \xmm~observation of \iras~suggests  the possibility 
that the absorber along the line of sight to the nucleus of \iras~is
Compton-thin. 
If this is the case, it implies a scenario completely different
from that reported for this Type 2 quasar so far.
It is worth stressing, however, that previous X-ray studies of \iras~inferred a
reflection-dominated nature of its 2--10 keV spectrum mainly
on the basis of poor signal-to-noise $\sim$15--50 keV data taken with the
non-imaging \beppo/PDS detector.
Future imaging spectroscopy of \iras~performed in the 
10--50 keV range, say with {\it Simbol-X} or {\it XEUS}, 
is therefore needed to make definitive progress in measuring the exact continuum
emission 
from the quasar and
constraining the column density of the nuclear absorber.
A deep \chandra~observation of \iras~would also be useful to
accurately quantify the strength of the \feka~emission line at 6.4 keV, which
is a proxy for the Compton thickness of the absorber.

\begin{acknowledgements}
We thak the referee, Dr. K. Iwasawa, for careful reading and for many
  useful comments that helped us to improve the manuscript.
We are grateful to G. Miniutti, R. Maiolino, S. Colafrancesco and G. Risaliti for helpful
discussions. Based on observations obtained with \xmm, an ESA science
  mission with instruments and contributions directly funded by ESA Member
  States and NASA. This research has made use of the NASA/IPAC Extragalactic 
  Database (NED) which is operated by the Jet Propulsion Laboratory,
  California Institute of Technology, under contract with the National
  Aeronautics and Space Administration. We  acknowledge financial contribution from
 contract ASI-INAF I/023/05/0.
\end{acknowledgements}
%%%%%%%%%%%%%%%%%%%%%%%%%%%%%%%%%%%%%%%%%%%%%%%%%%%%%%%%%%%%%%%%%%%%%%%%%%%%%%%%%%%

\end{document}